# Information fusion strategy integrating pre-trained language model and contrastive learning for materials knowledge mining

Yongqian Peng[+], Zhouran Zhang[+,*], Longhui Zhang, Fengyuan Zhao, Yahao Li, Yicong Ye[*], Shuxin Bai

+These authors contributed equally to this work. *Corresponding authors.

College of Aerospace Science and Engineering, National University of Defense Technology

Changsha, Hunan, China, 410073

## Abstract

Machine learning has revolutionized materials design, yet predicting complex properties like alloy ductility remains challenging due to the influence of processing conditions and microstructural features that resist quantification through traditional reductionist approaches. Here, we present an innovative information fusion architecture that integrates domain-specific texts from materials science literature with quantitative physical descriptors to overcome these limitations. Our framework employs MatSciBERT for advanced textual comprehension and incorporates contrastive learning to automatically extract implicit knowledge regarding processing parameters and microstructural characteristics. Through rigorous ablation studies and comparative experiments, the model demonstrates superior performance, achieving coefficient of determination ($R^2$) values of 0.849 and 0.680 on titanium alloy validation set and refractory multi-principal-element alloy test set. This systematic approach provides a holistic framework for property prediction in complex material systems where quantitative descriptors are incomplete and establishes a foundation for knowledge-guided materials design and informatics-driven materials discovery.

## Introduction

Artificial intelligence is revolutionizing the paradigm of materials development[1–4]. Over the past decade, machine learning-assisted materials design methodologies have achieved unprecedented breakthroughs in multiple domains, including crystal structure generation[5], phase stability analysis[6] and intrinsic properties prediction[7], demonstrating computational efficiency improvements of several orders of magnitude compared to conventional approaches. However, in stark contrast to computer vision[8] and natural language processing[9], materials science grapples with a distinctive small-sample predicament[3], where reliable experimental data remains scarce and prohibitively expensive to acquire. Although researchers have applied various mitigation strategies, including data augmentation[10,11], active learning[12,13], and transfer learning[14,15], yielding promising results in specific cases, the predominant AI4M (AI for Materials) approaches remain circumscribed by a reductionist perspective: simplifying materials properties into deterministic relationships with specific feature parameters and constructing such mappings through feature engineering and model training. This methodology has proven particularly effective in predicting intrinsic materials properties[16,17]. Its fundamental advantage lies in the selected features' capacity to capture the critical information governing target properties, thereby establishing explicit physical correlations between features and properties.

Nevertheless, many materials properties remain difficult to fully characterize through reductionist approaches, with the prediction of ductility in metallic materials serving as a typical example. In materials science, ductility is defined as a material's ability to undergo significant plastic deformation before fracture, typically quantified by elongation at fracture during tensile testing[18]. Ductility is not merely influenced by elemental composition and stoichiometry but is profoundly dependent on processing features (such as processing approaches, heat treatment recipes, and deformation processes)[18–21], the resulting microstructural characteristics (such as grain boundary features, secondary phase distribution, and elemental segregation)[22,23], and processing-related defects (such as porosity, shrinkage cavities, and dislocations)[24,25]. These non-intrinsic factors resist effective parameterization and model incorporation, thereby impeding the models' generalizability to novel processing routes or alloy systems. These factors typically modulate material properties through intricate interactions that resist simple numerical representation. This challenge is particularly pronounced in complex multi-principal-element alloy (MPEA) systems. Traditional empirical criteria (such as the K/G ratio, valence electron concentration, and atomic size difference) exhibit limited predictive capability [26,27], while machine learning models, despite their capacity for high-throughput screening of vast datasets, remain constrained by the comprehensiveness of feature selection[28,29], potentially resulting prediction errors in extrapolation scenarios.

Materials science literature presents a promising avenue for addressing these challenges. These texts not only encompass experimental data but also encapsulate authors' profound experimental insights and distilled materials science knowledge. Recent years have witnessed remarkable advances in natural language processing techniques for materials literature mining [30–32], particularly with the emergence of pre-trained large language model-based data extraction methods[33]. These approaches enable efficient structured data acquisition from semi-structured tables and unstructured text, thereby establishing a robust foundation for subsequent machine learning model construction[34–36]. For instance, Dagdelen et al.[34] achieved precise automated extraction of structured materials data by developing specialized prompting templates based on GPT and LLaMA models. The extracted structured data can also be presented in the form of a knowledge graph and utilized to accomplish downstream tasks related to material properties[37,38].

However, the knowledge embedded within materials literature transcends directly extractable data, encompassing vast domains of difficult-to-structure expertise and empirical insights. Consequently, researchers have begun exploring more sophisticated text comprehension methodologies. Early endeavors employed neural network models like Word2Vec[39] to transform unstructured text into high-dimensional vectors for end-to-end representation learning, achieving preliminary success in domains such as ferroelectric materials. Subsequently, the high-dimensional vectors containing knowledge in the domain of material science were utilized to predict material properties using an LSTM-based neural network[40]. Recently, with the evolution of Transformer architecture, pre-trained models in materials science have achieved significant breakthroughs. Models such as MatSciBERT[41], MaterialsBERT[42] have substantially enhanced semantic comprehension of materials-specific texts through pre-training on vast materials literature. Even some more specialized pre-trained language models have emerged in niche domains, such as PolyBERT[43] and SteelBERT[44]. These models demonstrate sophisticated understanding of complex relationships between chemical composition, structural characteristics, and properties, pioneering new pathways for mining implicit knowledge from texts. The significance of this text vectorization approach lies in its ability to represent features (such as processing details and microstructural characteristics) that resist quantification from a reductionist perspective, enabling more comprehensive extraction and utilization of knowledge embedded in texts. Nevertheless, the integration of this rich textual information with materials knowledge, particularly in predicting complex properties like materials ductility remains a critical scientific challenge awaiting resolution.

In this study, we present an innovative information fusion architecture that integrates domain-specific texts from materials science literature with quantitative physical descriptors, thereby overcoming the limitations of traditional reductionist approaches that struggle to fully quantify all microstructural influences. This integration significantly enhances the model's capacity for information representation and generalization. Our framework uses MatSciBERT for advanced textual comprehension and incorporates contrastive learning strategy, which allows automatic extraction of implicit knowledge from the literature, such as processing parameters and microstructural characteristics. These elements supplement and enrich the descriptors required for property prediction and enable the inference and prediction of critical, otherwise unobservable variables. By analyzing latent relationships among varied textual entries, the model is able to address challenges arising from high experimental costs or inaccessible material parameters. To demonstrate the effectiveness of this approach, we curated a specialized corpus containing 313 titanium alloy entries and 55 refractory multi-principal-element alloy (RMPEA) entries, each including ductility data and influencing text descriptions such as processing details and microstructural characteristics. Rigorous multi-level comparative analyses were performed to validate the performance of the numerical and textual information fusion strategy, including assessments of transferability to other alloy systems. By applying the resulting model to the compositional space of the Ti-V-Zr-Nb-Hf-Ta six-component system, we further explored the potential for alloy design and knowledge discovery without relying on prior microstructure information.

# Results

## Dataset Construction

The dataset is composed of four different kinds of materials information. The first kind (①) is 18 materials attributes (mean, mismatch, and variance of atomic radius, electronegativity, valence electron concentration, Young's modulus, shear modulus, and bulk modulus) that have been established elsewhere as useful descriptors[26,45,46]. These materials attributes are likely to enhance the physical information missing from the literature. It is worth noting that we do not directly use the numerical features of these material attributes, but rather transformed the numerical features into text input via a template, as shown in the Fig. 1a (the reasons will be discussed later). Since processing and microstructure significantly affect ductility, the other two kinds of materials information are directly extracted from publicly available literature and in the form of text, including ②processing details and ③compositional and microstructural characteristics. As shown in Fig. 1b and Fig. 1c, the compositional and microstructural information are often found in the results and discussion section and they are grouped together primarily because microstructural information is often accompanied by compositional descriptions. The last kind of information is the alloy name ④, marked red in the texts which can provide important material composition information, which plays a crucial role when different alloys have similar processing conditions but different ductility. The fracture elongation rate, namely ductility, is extracted and used as the target variable for predictions. The text extraction process is depicted in Fig. 2a, with a detailed construction process to be elaborated in the **Materials and Methods** section.

It is important to note that during the text cleaning step, we preserve descriptions of both alloy preparation methods and post thermal-mechanical treatment processes. For example, we retain sentences such as "The test powders are dried using a vacuum oven for 12 h at a temperature of 100°C" and "All samples with 12 mm thickness were solution treated at 950°C for 1 h." Additionally, for the composition and microstructure texts, apart from texts providing direct descriptions of the microstructure, we also keep statements reflecting the influence of composition on microstructure and mechanical properties, such as "The smaller the grain size, the higher the level of performance. The grain size of Ti-xFe alloys decreases as the Fe content increases.”. To avoid negative effects from sparse data associated with certain specialized processing techniques, entries including processing methods that appear only once in the dataset are removed. After cleaning, 313 titanium alloy corpora are reduced to 226 final entries (with 80% allocated to the training set and 20% to the validation set). As mentioned previously, for cases where alloys have consistent processing but different microstructures and ductility due to variations in alloy composition (a relatively common scenario), we extract only those textual descriptions with distinct microstructural characteristics and included alloy names as word vectors to emphasize compositional differences. It is worth noting that among 55 RMPEA entries, a dataset of 30 entries is established to validate the model’s reliability and generalization capability on RMPEAs. Additionally, to further explore the model’s performance and to determine whether it is possible to learn the underlying patterns influencing ductility, we retain 25 entries whose processing methods and alloy systems are consistent, but the content of the constituent elements varies, resulting in different microstructures and ductility values.

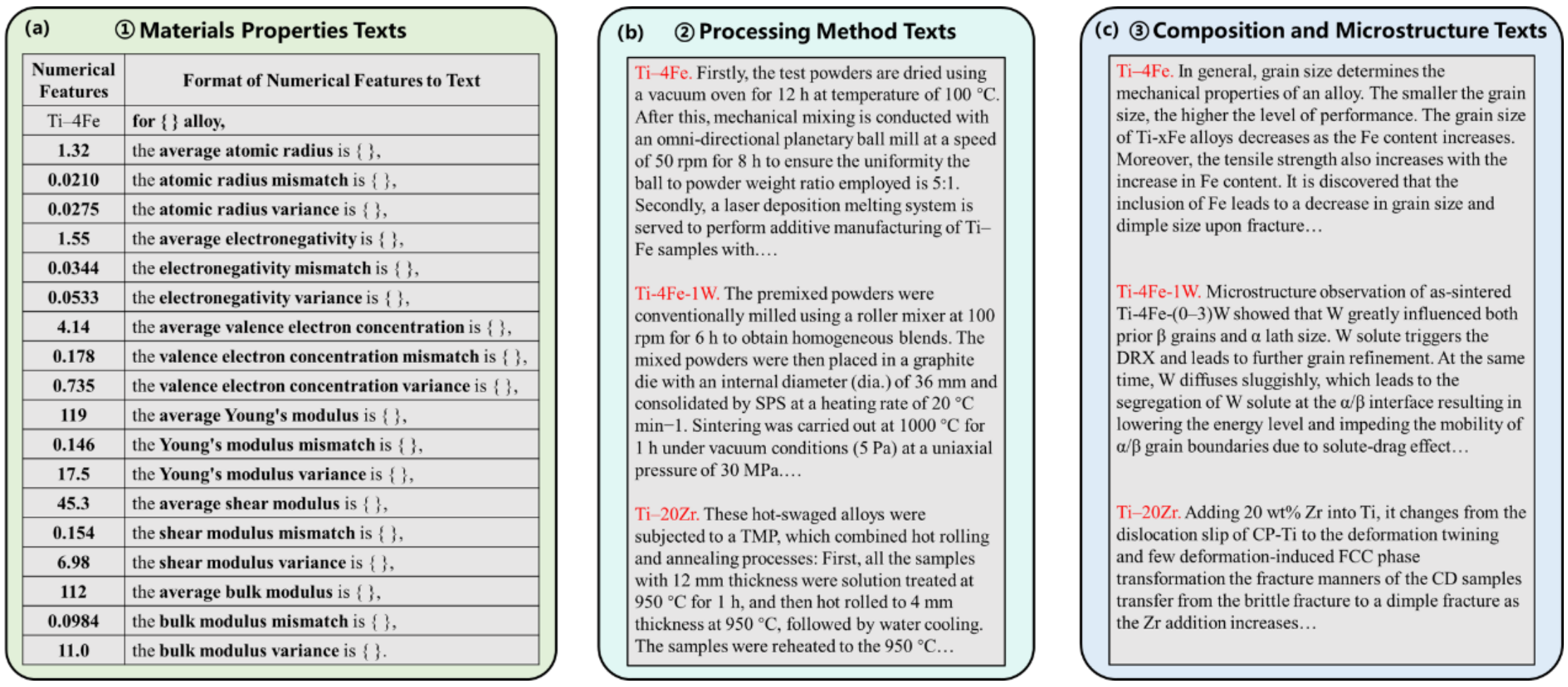

**Fig. 1| Examples of model inputs.** (a) Materials properties texts, (b) Processing method texts, and (c) Composition and microstructural texts.

## Information fusion model

Fig. 2a shows the process of corpus extraction and cleaning as well as text tokenization. The text data, after being cleaned and formatted, is encoded and used as input to the model. Fig. 2b illustrates the architecture of the information fusion model. MatSciBERT serves as the text embedding extractor to obtain high-dimensional representations of materials science texts.

These high-dimensional representations are further processed to construct semantic vectors capable of describing the elongation of alloys. Subsequently, supervised learning is performed using these semantic vectors as input features and elongation values as labels. Upon completion of training, the model can be deployed to predict the elongation of alloys with the information of ①+②+③+④. All implementation details of the model are provided in the **Materials and Methods** section.

The loss curve of the training process is depicted in Fig. 3a, where the mean squared error (MSE) gradually decreases as the training epochs increase, ultimately converging to a small value. This observation indicates that the model architecture designed for text input is reasonable and that the training is effective. Fig. 3b demonstrates the model's fitting performance on the titanium alloy training and validation sets, as well as the RMPEA test set. The coefficient of determination ($R^2$) values for the titanium alloy training and validation sets reach 0.899 and 0.845, respectively. Compared to traditional machine learning models (SVM, RF and MLP), as shown in Fig. 3b (results in detail is given in Table S1 in **Supplementary Information**), the $R^2$ values exhibit a significant improvement. This enhancement is attributed to two key factors. Firstly, the information fusion mechanism integrates textual information from material properties, microstructure, and processing methods, resulting in a more comprehensive and richer feature description. Secondly, the information fusion model benefits from MatSciBERT's training on a large corpus of texts, enabling the model to grasp general knowledge within the materials domain and capture complex relationships and influencing factors that traditional machine learning models, which utilize only single numeric features as input, fail to grasp.

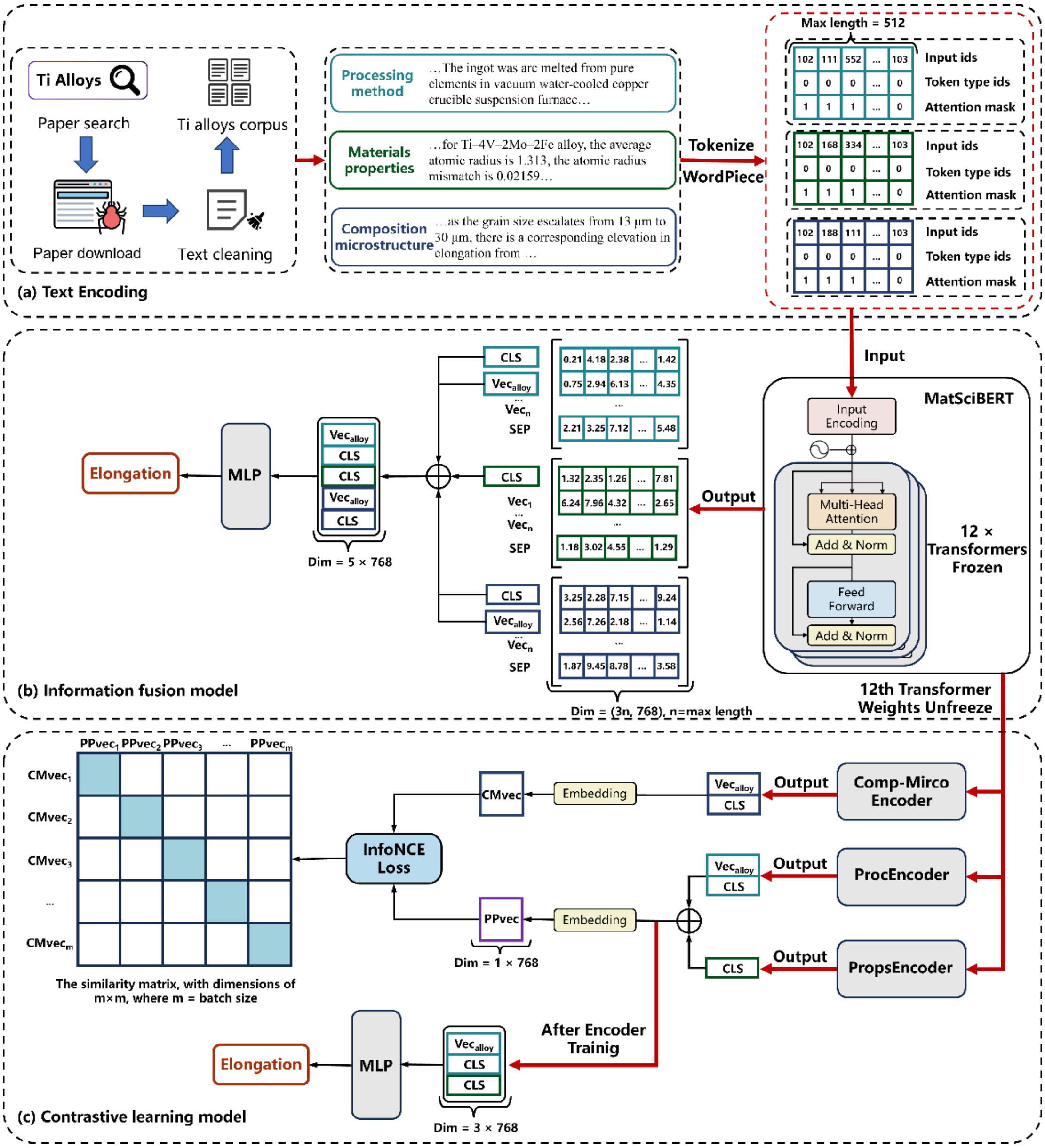

**Fig. 2| Overview of the approach**. (a) Text Encoding. Published titanium alloy and refractory multi-principal-element alloy papers were collected, cleaned, and tokenized (max length 512, WordPiece algorithm) for sections on processing methods, material properties, composition, and microstructure. These were encoded into input ids, token type ids, and attention masks as inputs for MatSciBERT. (b) Text embeddings for processing methods and composition/microstructure were obtained by concatenating CLS vectors with alloy word vectors, while the property embedding used only the CLS vector. All three embeddings were concatenated and fed into a fully connected neural network (MLP) to predict fracture elongation. (c) Contrastive learning model. We trained separate encoders for composition/microstructure, processing methods, and material properties based on MatSciBERT using contrastive learning. By modeling the relationships among these embeddings, the system can predict alloy fracture elongation from limited input information.

Given the considerable differences in composition, structure, and properties among various alloy systems, achieving good generalization for cross-system alloy performance prediction has historically been challenging using previous methods that rely solely on numeric features as input[3,47]. In our work, the information fusion model still achieved an $R^2$ score of 0.680 on the RMPEA test dataset. This improvement stems from two complementary mechanisms: the supervised fine-tuning process optimizes the model's parameters for specific tasks, enabling it to capture task-specific features, while the pre-trained model's ability to learn general language representations from extensive generic text data establishes connections and distinctions between different alloy systems. Therefore, we believe that the approach of utilizing a pre-trained language model followed by fine-tuning is one of the effective solutions to the issue of inadequate generalization in cross-alloy system performance predictions.

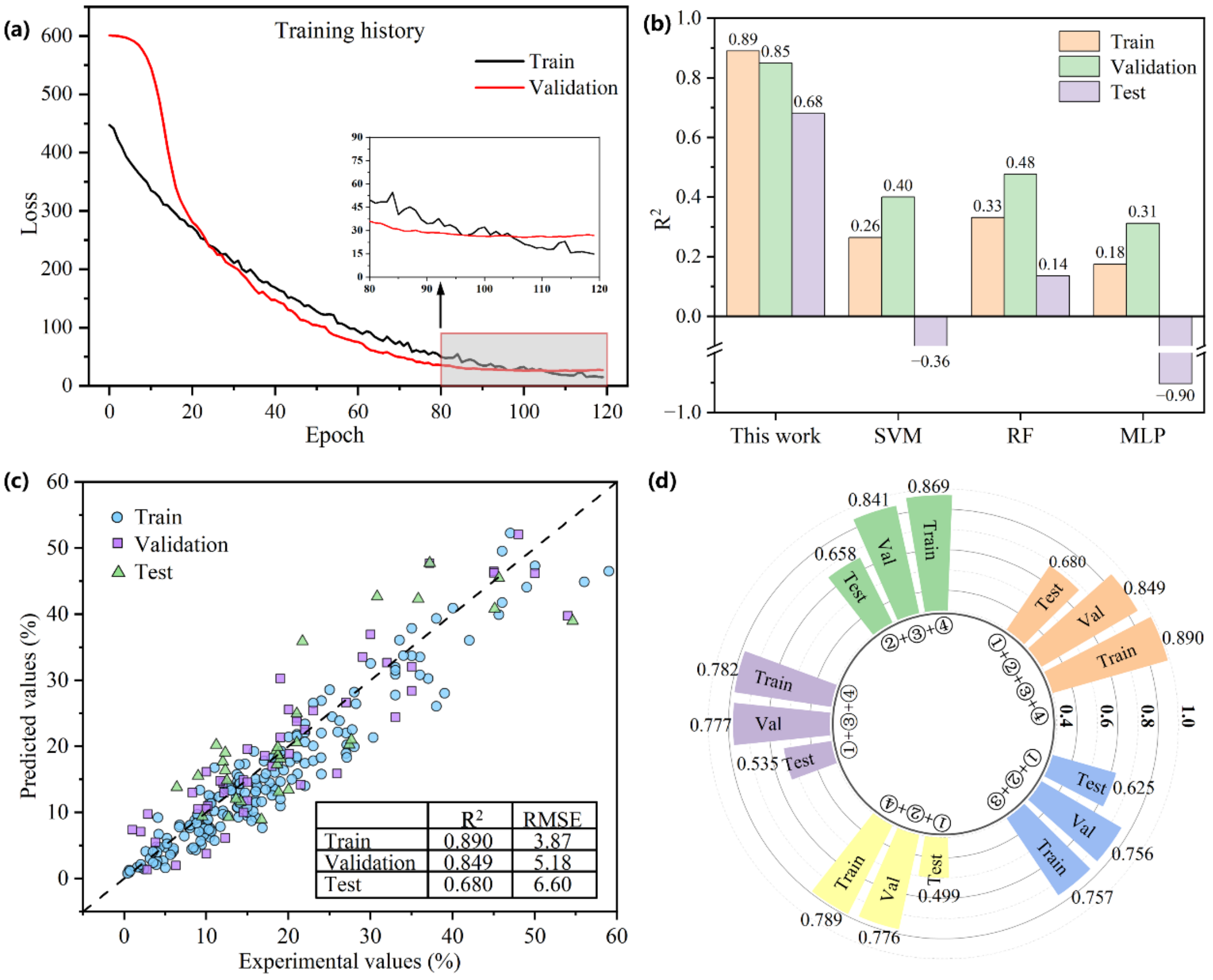

**Fig. 3| Model performance and ablation study results.** (a) Training and validation loss curves (MSE) for the information fusion model, showing low prediction error and minimal overfitting. (b) Comparison of the information fusion model with three traditional numerical feature-based machine learning models on titanium alloy (train/validation) and refractory multi-principal-element alloy (test) datasets. (c) Parity plot of predicted vs. experimental elongation values for the information

fusion model; most large deviations occur in the high-elongation region. (d) Ablation study results, where ①②③④ represent materials properties texts, processing method texts, composition and microstructural texts and alloy name, respectively, each component's contribution to model performance is evaluated.

## Ablation studies

To assess the significance of textual features related to material properties, processing methods, and composition and microstructure, as well as the alloy word vector features in predicting alloy fracture elongation, we conducted ablation experiments. By sequentially removing each category of features, we observed changes in model performance to determine the contribution of each feature type within the model. In all ablation experiments, the hyperparameters of the model remain consistent with those of the information fusion model. The results are illustrated in Fig. 3d. Following the ablation of material properties text, the $R^2$ values for the training, validation, and test datasets experience only slight declines (0.021, 0.008, and 0.022, respectively), indicating that the removal of material properties text did not significantly impact model performance. Although the [CLS] token is used as a high-dimensional representation of the text, achieving notable improvement of performance in predicting fracture elongation, this text format presents inherent limitations in detecting fine-grained variations in material properties. Specifically, material properties text employs a standardized sentence structure where only key numerical values are modified while maintaining identical linguistic patterns. When the semantic information of such sentences is condensed into the [CLS] token representation, the model exhibits reduced sensitivity to these subtle numerical variations. Consequently, the [CLS] token demonstrates limited capability in distinguishing informational differences between sentences that differ primarily in numerical values, thereby constraining the contribution of material properties text to model performance. Nevertheless, material properties text provides valuable contextual information that complements other textual features, enhancing the model’s overall comprehension of ductility behavior in materials.

In contrast, after removing the text regarding processing methods, the $R^2$ values decreased by 0.108, 0.072, and 0.145 on the training, validation and test datasets respectively. Similarly, ablation of composition and microstructure text caused reductions of 0.101, 0.073, and 0.181 in the respective datasets. This indicates that both ②processing and ③composition and microstructure text play crucial roles in the model, providing essential valuable information that enhances the model's comprehensive understanding and prediction of alloy fracture elongation. The substantial drop in the $R^2$ scores upon removal of these text categories demonstrates that the model's predictive capability is heavily reliant on this information. This finding further validates the advantages of the information fusion model in integrating multi-source data, uncovering complex relationships, and enhancing predictive performance.

## Multi-level comparison

So far, we have constructed an information fusion model using MatSciBERT as the text embedding extractor to predict the fracture elongation of titanium alloys. To highlight the superiority of our proposed information fusion strategy, a multi-level comparative experimental framework (shown in Fig. 4) was employed to systematically evaluate the benefits of various information fusion strategies in materials property prediction. First, three baseline models (SVM, RF and MLP) utilizing only 18 physical numerical features of materials—without any textual information—were used as the simplest control. Second, a direct fusion strategy, wherein the numerical features were concatenated with text features extracted by MatSciBERT and then fed into a neural network, allowed for assessment of the most straightforward multimodal integration. Subsequently, a hierarchical transformation strategy was adopted: the MatSciBERT-derived textual features (dimension 768×4) and the 18 physical numerical features were each independently transformed via dedicated fully connected layers, and the outputs were concatenated into a composite feature vector for further processing by the neural network, thereby enhancing inter-modal cooperation and representation capacity. In the fourth approach, GloVe[48] word embeddings were used instead of MatSciBERT for textual encoding, following the same hierarchical fusion process, to evaluate the impact of different text representations on model performance. Finally, a question-answering model based on DeepSeek-V3[49] was constructed, combining natural language understanding with materials data analysis to explore the potential of large language models for knowledge generalization and transfer in materials science. In this approach, the three types of textual information were used as natural language prompts for a Q&A task predicting target alloy ductility. This progressive model design enables a comprehensive comparison of multimodal fusion and semantic modeling methods in the context of materials prediction tasks. Detailed results are given in the **Supplementary Information.**

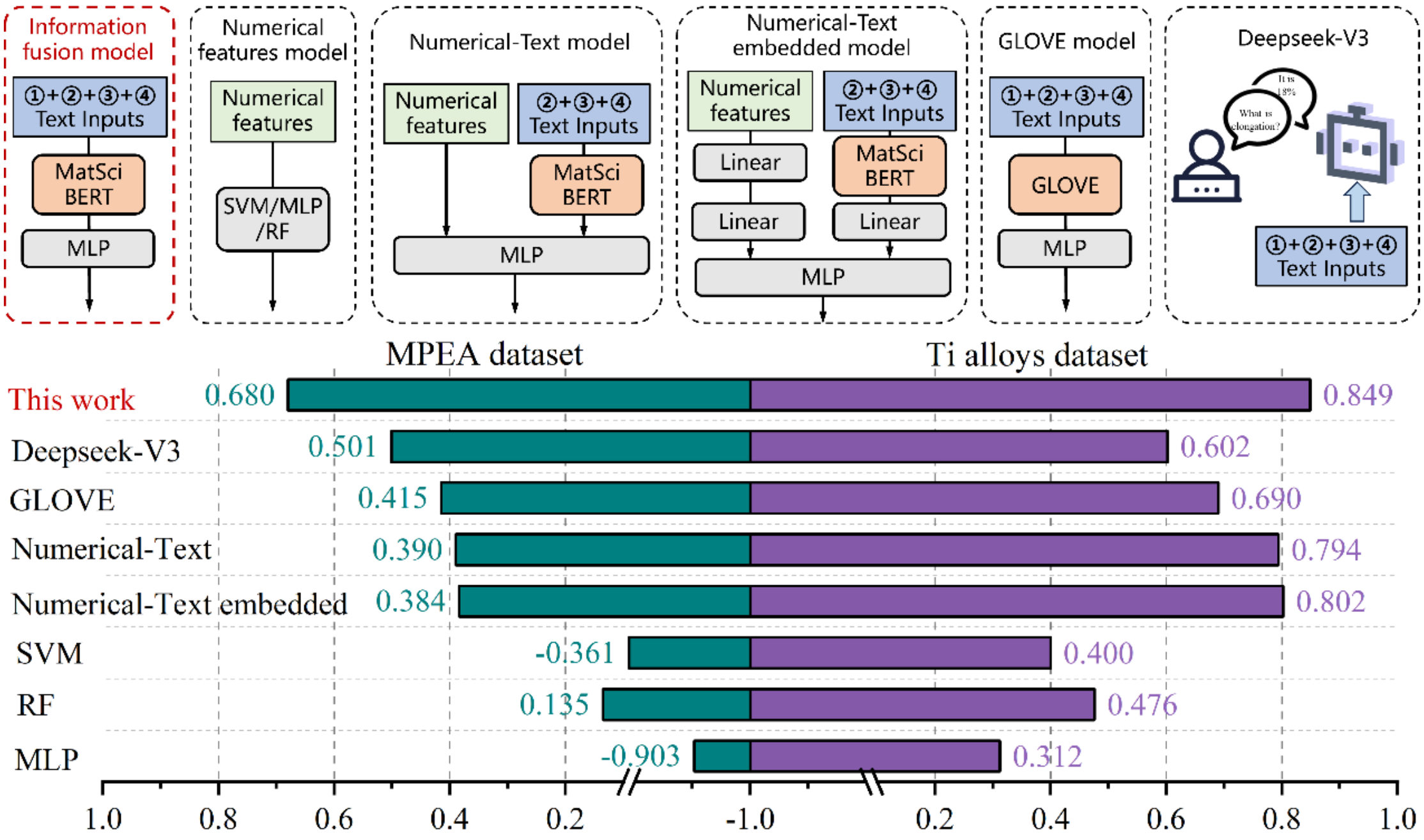

**Fig. 4| The multi-level comparative experimental results.** Machine learning models constructed using different strategies for predicting alloy ductility, where boxed text provides schematic representations of each strategy. From left to right are: information fusion model based on MatSciBERT, baseline models utilizing only numerical features without any textual information, the numerical-text model (a direct fusion strategy), the numerical-text embedded model (in which features are independently transformed and then concatenated into a composite feature vector), the GloVe-based information fusion model, and the question-answering model based on DeepSeek-V3. The bar chart at the bottom of the figure displays the performance ($R^2$) of each strategy in both titanium alloys and RMPEAs datasets.

As shown in Fig.4, baseline models utilizing only numerical features exhibit the lowest predictive performance, reflecting the limitations of solely numerical, reductionist methods. The numerical-text model using a direct fusion strategy, achieves $R^2$ values on the Ti alloy dataset close to those of the information fusion model (0.794), but demonstrates significantly lower performance on the RMPEA dataset (0.390), highlighting the challenges of directly merging different types of data (Fig. S1a). Notably, unlike the information fusion model, we employed a larger learning rate for the numerical-text model to achieve optimal training results. This may be due to the discrepancy in parameter counts between the fully connected network designed for numerical feature extraction and the MatSciBERT model used for extracting text embeddings, leading to differences in the depth of information captured and thus complicating the fusion process. The numerical-text embedded model, which applies two separate linear layers to transform the extracted numerical and textual features before integration, performs similarly to the numerical-text model (Fig. S1b), achieving an $R^2$ score of 0.802 on the Ti alloys dataset while exhibiting a significant decline to 0.384 on the RMPEA dataset. These findings suggest two key insights. First, simple linear transformation is insufficient to bridge the gap arising from network architecture mismatches, thus failing to adequately capture and integrate heterogeneous feature information. Second, we suggest that the complexity and depth of feature transformation may be inadequate to address the inherent complexity of the RMPEA dataset, which further underscores the superiority of the information fusion approach based on MatSciBERT for readily achieving alignment across disparate information modalities.

The GloVe-based information fusion model, with text feature extractor replaced by a non-attention-based GloVe language model, achieves moderate performance (e.g., $R^2$ values of 0.690 for Ti alloy and 0.415 for RMPEA datasets), considerably less than the MatSciBERT-based model (Fig. S2). Despite the abundance of information, the choice of an effective text embedding technique is critical. The GloVe model, pre-trained on general text rather than materials-related corpora, is unable to adequately represent essential materials science concepts, thus limiting its predictive accuracy and generalizability. Although, the question-answering model based on DeepSeek-V3 achieves moderate $R^2$ values (e.g., 0.552 for Ti alloys and 0.482 for RMPEAs), its performance is still significantly inferior to that of the fine-tuned MatSciBERT-based information fusion model. The Q&A records in detail is given in Table S2. Notably, DeepSeek-V3 often generates flexible and insightful responses, sometimes referencing processing or microstructure details, but its accuracy greatly depended on the explicitness of the numerical information within the text. It tends to excel at summarizing relationships present in the data but struggled to infer materials science knowledge that is not overtly presented, revealing the limitations of general-domain pre-training without materials-specific fine-tuning. Nevertheless, this flexible response style suggests a promising direction for future large language models in scientific knowledge transfer and understanding[50,51].

To briefly sum up, this multi-level comparison highlights that the MatSciBERT-based information fusion model, finely tuned for domain-specific tasks, consistently outperform general language models and non-domain-specific fusion strategies. The results underscore the importance of both domain-adapted text embedding and targeted supervised fine-tuning for achieving robust performance and generalizability in materials property prediction.

## Contrastive learning model

Ablation experiments indicate that text related to composition and microstructure is crucial for accurately predicting the fracture elongation of alloys. However, for unknown alloys, the effects of composition and microstructural characteristics on performance are typically unknown prior to experimentation. Consequently, this information fusion approach cannot be directly used to predict the ductility of unknown alloys. To bridge this gap, we developed a contrastive learning strategy whose core idea is to partition data samples into positive and negative sample pairs. By optimizing a contrastive loss function, the approach brings the feature representations of positive pairs closer together while pushing those of negative pairs farther apart, thereby learning discriminative feature representations that better capture the intrinsic structure and semantic information of the data. In this work, as shown in Fig. 1c we consider a material's set of information—materials properties, processing methods, and composition and microstructure—as positive sample pairs, while treating information from different materials as negative pairs (for example, the processing method of Alloy A and the microstructure of Alloy B form a negative pair). We construct three encoders and train them using contrastive learning to extract semantic vectors for material property texts, processing method texts, and composition and microstructure texts, respectively. The design of these encoders aims to capture the underlying semantic relationships within the texts, thereby revealing the structure-property linkages among material properties, processing methods, composition variations, and microstructural features. This enables the model to accurately predict alloy performance even without explicit input on composition and microstructure. Due to limited data availability, training a contrastive learning model from scratch is impractical. Moreover, since MatSciBERT already exhibits strong feature extraction capabilities for materials science texts, we fine-tune part of its parameters to construct the three encoders. The detailed model architecture and training procedures are provided in the **Materials and Methods** section.

Fig. 5a presents the loss curve for the contrastive learning model during training. It can be observed that the loss values for both the training and validation sets ultimately converge to approximately 0.3, indicating that the trained model effectively captures the relationships between the alloys' material properties, processing methods, and composition and microstructure. Following contrastive learning training, the model is applied to the elongation prediction task using an identical framework to the information fusion model, except that textual inputs describing alloy composition and microstructure are omitted. The predictive performance of the model for elongation in Ti alloys and RMPEAs is illustrated in Fig. 5b. The contrastive learning model achieves comparable performance to the information fusion model on both training and test sets, with $R^2$ values reaching 0.886 and 0.673, respectively. These results demonstrate the effectiveness of the contrastive learning strategy.

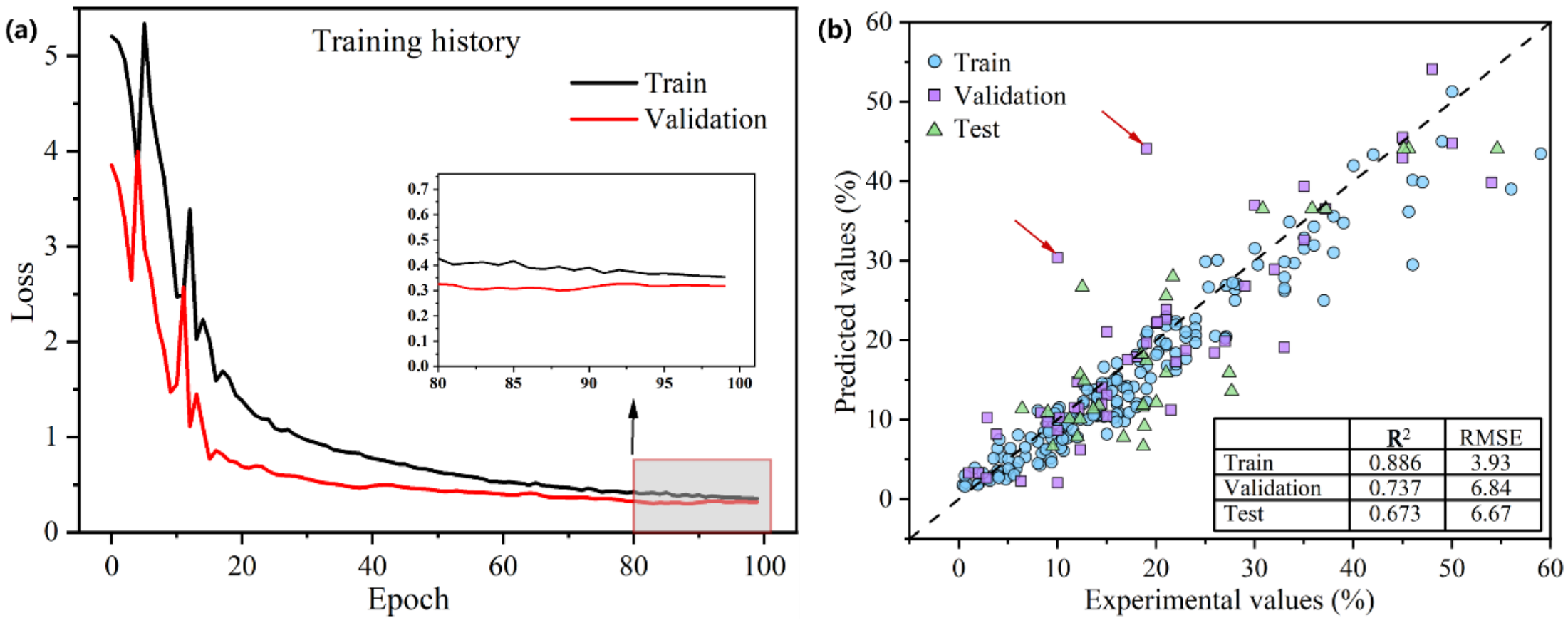

**Fig. 5| Contrastive learning results.** (a) Training loss curves for the three encoders. The custom-designed symmetric loss function simultaneously accounts for bidirectional learning: from material properties and processing methods to composition and microstructure, and vice versa—from composition and microstructure to material properties and processing methods. (b) Scatter plot of the data points predicted by the contrastive learning model.

In fact, similar to the information fusion model, our contrastive learning model can also be viewed as a form of information fusion strategy. In the information fusion model, we directly concatenate the text features to obtain a high-dimensional semantic vector encompassing information from the three categories of features. In the contrastive learning model, we fuse information by comparing similarities to achieve a high-dimensional semantic vector that also contains information from the three categories of features. However, a significant advantage of the contrastive learning strategy over the information fusion model is that, after training on a certain number of samples, the model establishes connections between processing methods, composition changes, microstructure, and material properties. Thus, during the testing phase, the extracted high-dimensional semantic vectors for material properties and processing methods inherently contain information about the microstructure. This enables the model to enhance its predictive capability even in the absence of textual features related to composition changes and microstructure, thereby facilitating predictions of alloy ductility. Despite the model's overall strong performance on the training set and the RMPEA test set, its performance on the Ti alloy validation set is lower than that of the information fusion model. Careful analysis of the data within the validation set reveals that the $R^2$ discrepancy can be attributed to two outliers marked in the figure. This may be due to the model establishing incorrect relationships between the alloy properties, processing methods, and microstructural characteristics during contrastive learning, which led to cumulative errors during the transition from the contrastive learning training phase to the fracture elongation prediction phase.

## Contrastive learning model validation

In materials science, guided by the principle of controlled experiments, it is common practice to systematically vary the content of a single element in order to investigate its effect on microstructure and mechanical properties. This approach is also frequently employed in studies on the ductility of RMPEA. Accordingly, in such cases, our contrastive learning strategy is required to rely solely on alloy composition information to accurately predict the ductility of the alloys. This is an exceedingly challenging task, which is also highly susceptible to overfitting if the model has not acquired underlying knowledge of materials science. For this purpose, we adopt a posterior validation strategy using experimentally reported data from the literature, specifically utilizing the previously mentioned reserved set of 25 samples. We applied the contrastive learning model to directly predict the ductility of these 25 RMPEAs, with processing methods described as follows: "The XX alloy was prepared by arc-melting using a mixture of pure raw metals (purity ≥ 99.95%), and was remelted at least five times to promote the chemical homogeneity." The experimental values, the contrastive learning predictions for these completely unseen data samples, and the errors calculated as the difference between experimental and predicted values are shown in Table 1. It is worth noting that we also evaluated a second processing text template: "The XX alloy was synthesized by arc melting a blend of high-purity metals (purity ≥ 99.95%) and subsequently remelted at least five times to ensure chemical homogeneity." The resulting ductility predictions were found to be very similar between the two templates.

**Table. 1|** The experimental values and contrastive learning predicted values for the completely unseen data samples, as well as the errors calculated by experimental value minus predicted value.

| Alloy | Experimental Elongation (%) | Predicted Elongation (%) | Error (%) | Ref. |
|---|---|---|---|---|
| TiNbTa | 17 | 15.9 | 2.6 | [54] |
| $Ti_{57.2}Zr_{18.3}Nb_{24.5}$ | 14 | 9.4 | 4.6 | [55] |
| $Ti_{54.1}Zr_{23.9}Nb_{22.0}$ | 12 | 8.1 | 3.9 | [55] |
| $Ti_{52.6}Zr_{26.7}Nb_{20.7}$ | 11 | 7.8 | 3.2 | [55] |
| $Ti_{50.3}Zr_{31.1}Nb_{18.6}$ | 9.5 | 6.8 | 3.1 | [55] |
| TiZrHf | 13.6 | 12.5 | 1.1 | [53] |
| $Ti_{40}Zr_{25}Nb_{25}Ta_{10}$ | 18 | 18.8 | 0.8 | [56] |
| $Ti_{45}Zr_{25}Nb_{25}Ta_{5}$ | 22 | 20 | 2 | [56] |
| $Ti_{31.67}Zr_{31.67}Nb_{31.66}Ta_{5}$ | 25 | 23.7 | 1.3 | [53] |
| $Ti_{35}Zr_{35}Nb_{25}Ta_{5}$ | 21 | 20.2 | 0.8 | [53] |
| $Ti_{45}Zr_{45}Nb_{5}Ta_{5}$ | 20.5 | 19 | 1.5 | [53] |
| $Ti_{21.67}Zr_{21.67}Nb_{21.66}Ta_{35}$ | 12.7 | 18.6 | **-5.9** | [53] |
| $TiZrHfTa_{0.3}$ | 13.96 | 18.7 | -4.74 | [53] |
| $TiZrHfTa_{0.4}$ | 30.26 | 19 | **11.26** | [53] |
| $TiZrHfTa_{0.5}$ | 29 | 16.5 | **12.5** | [53] |
| $TiZrHfTa_{0.6}$ | 21.73 | 16.1 | 4.83 | [53] |
| $TiZrHfTa_{0.7}$ | 14.31 | 14.6 | -0.29 | [53] |
| $TiZrHfTa_{0.8}$ | 4.63 | 13.7 | **-9.07** | [53] |
| TiZrNbHf | 8/12 | 10.1 | -2.1/1.9 | [53,57] |
| $TiZrNb_{0.2}HfTa_{0.2}$ | 32.5 | 21.3 | **11.2** | [53] |
| $TiZrNb_{0.4}HfTa_{0.4}$ | 10 | 19 | **-9** | [53] |
| $TiZrNb_{0.5}HfTa_{0.5}$ | 13 | 17.6 | -4.6 | [53] |
| $TiZrNb_{0.6}HfTa_{0.6}$ | 19 | 18 | 1 | [53] |
| $TiZrNb_{0.8}HfTa_{0.8}$ | 18 | 16.2 | 1.8 | [53] |
| TiZrNbHfTa | 13.5 | 13 | 0.5 | [54] |

Note: The experimental processes of all alloys and the process texts used as input for contrastive learning are all in the as-cast condition.

As shown in the Table 1, except for the six bolded data points where the prediction error exceeds 5%, the remaining errors are relatively small. This is a very encouraging result for the prediction of ductility, considering that, even when using as-cast processing conditions, differences of up to 5% in reported ductility values for the same composition are common across different research groups. Nonetheless, the results also indicate that the model's ability to capture subtle variations in alloy composition still requires further improvement. For example, in the $TiZrHfTa_x$ system[52], although the model successfully captures the non-monotonic trend in ductility, namely ductility increases and then decreases with increasing Ta content, there is still a considerable difference between the predicted maximum ($TiZrHfTa_{0.4}$) and minimum ($TiZrHfTa_{0.8}$) values. It is well known that such behavior is mainly attributed to phase transformation induced plasticity of metastable phases during deformation[52]. The model's predictions in this regard appear to be relatively conservative. This issue is even more pronounced in the $TiZrNb_xHfTa_x$ system[53]; when both Nb and Ta contents vary simultaneously, the model is unable to accurately capture the trend of ductility changes.

Moreover, it is important to note that we are now able to predict the ductility of alloys without prior knowledge of their microstructures. Therefore, we applied the aforementioned as-cast processing conditions mentioned above to perform ductility predictions within the Ti-V-Zr-Nb-Hf-Ta six-component compositional space. Specifically, we predict the ductility of ternary, quaternary, and quinary alloys. Since our fundamental dataset is based on titanium alloys, we consider only those RMPEAs containing Ti, which results in a total of 10 ternary, 10 quaternary, and 5 quinary systems being selected. As our primary focus is on RMPEAs, the compositions are set within the range of 5-50 at% at 5 at% intervals. Here, we use the Ti-V-Ta ternary alloy (Fig. 6a) and the Ti-Zr-Nb-Ta quaternary alloy (Fig. 6b) as examples to illustrate the influence of compositional variations on ductility. The color bar representing ductility values is shown on the right side. As seen from Fig. 6a and 6b, the relationship between compositional changes and ductility is highly complex; the ductility value does not change monotonically with the variation of a single element. This indicates that the mechanical properties of MPEAs are subject to intricate interactions among constituent elements, which is markedly different from the property tuning rules for conventional single-principal-element alloys. This also indirectly demonstrates that the empirical criteria are not applicable in such scenarios involving cross-material systems and subtle variations in elemental composition. From Fig. 6c, it can be observed that in ternary alloys, the Ti-Nb-Ta and Ti-V-Ta systems exhibit the best ductility, while Ti-Zr-Ta, Ti-Zr-Nb, and Ti-Zr-Hf show poorer ductility, which agrees well with the prior report. In quaternary alloys, Ti-V-Nb-Ta is identified as potentially having the best ductility, whereas Ti-Zr-Nb-Hf exhibits relatively lower ductility. The differences in ductility among quinary alloys are small. Since the focus of this study is on developing an information fusion strategy for the challenging task of accurately predicting ductility from a reductionist perspective, we experimentally validated three representative alloys (marked in Fig. 6a by red circle) from the Ti-V-Ta ternary system: $Ti_{30}V_{40}Ta_{30}$ (predicted to have the highest ductility), $Ti_{30}V_{30}Ta_{40}$ and $Ti_{33}V_{33}Ta_{34}$ as comparative references. The model predictions for tensile elongation were 23.46%, 18.4% and 21.5%, respectively, while the experimental values were 24%, 12.5% and 22.5% as shown in Fig. 6d, respectively.

The experimental validation provides several important insights regarding the model's predictive capabilities and limitations. While the model demonstrates reasonable accuracy for the high-ductility alloy $Ti_{30}V_{40}Ta_{30}$, it exhibits large deviation for the lower-ductility composition $Ti_{33}V_{33}Ta_{34}$. The observed inaccuracy in predicting lower ductility values can be attributed to several factors. First, the corpus used for training the contrastive learning model could benefit from enrichment, particularly in the domain of processing-microstructure-property relationships specific to RMPEAs, which would enhance the model's predictive accuracy across the entire ductility spectrum. Second, a significant corpus bias likely exists in the available literature, as researchers tend to preferentially report RMPEAs with favorable ductility properties (as shown in Table 1). Consequently, a substantial amount of data regarding compositions and processing conditions that yield poor ductility remains unpublished, creating an imbalanced dataset that skews the model's learning toward high-performance scenarios, thereby affecting its overall robustness in covering the complete property space of RMPEAs. Nevertheless, predictions across the entire compositional space can still provide researchers with a preliminary framework for alloy design.

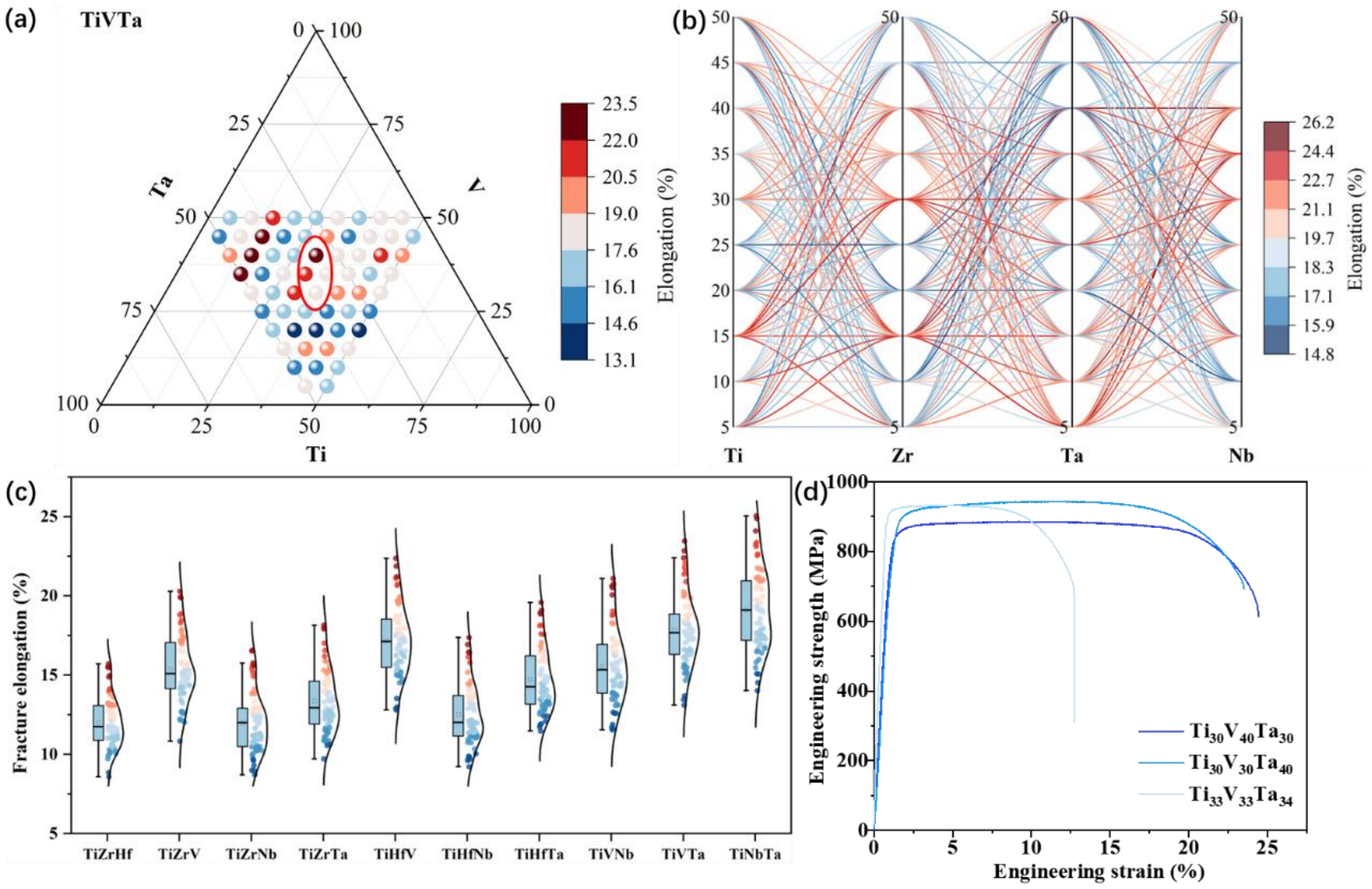

**Fig. 6| Compositional effects on RMPEA ductility.** Illustrative examples showing the influence of compositional variations on ductility for (a) Ti-V-Ta ternary and (b) Ti-Zr-Nb-Ta quaternary alloys. Ductility values are indicated by the color scale on the right. (c) Ductility comparison among different ternary RMPEAs, and (d) Engineering strength-strain curves for $Ti_{30}V_{40}Ta_{30}$, $Ti_{30}V_{30}Ta_{40}$ and $Ti_{33}V_{33}Ta_{34}$.

# Discussion

This work introduces an innovative information fusion architecture that combines advanced text comprehension with physical feature descriptors, representing a significant step forward in the field of materials informatics. Unlike traditional approaches that rely solely on hand-curated structured physical descriptors or purely textual analysis, our framework organically integrates domain-specific terminology from materials science literature (knowledge embeddings based on MatSciBERT) with quantitative physical features. This hybrid architecture greatly enhances the information representation and generalization capabilities of predictive models for materials properties.

A key contribution of this approach is its ability to overcome the inherent limitations of reductionist descriptor-based models, which struggle to quantitatively capture all microstructural influences on material behavior. By mining implicit knowledge within materials literature, including data on processing techniques and microstructural characteristics, our method systematically supplements and enriches conventional feature descriptions. This provides a more holistic and system-level foundation for complex property prediction tasks, especially in scenarios where explicit descriptors are incomplete or difficult to obtain.

The introduction of contrastive learning further enhances the model's unsupervised reasoning capabilities. Through comparative analysis of relationships among diverse textual entries, the framework enables inference and prediction of critical, otherwise unobservable variables such as unknown microstructural details. This capability addresses common challenges in experimental research, such as high data acquisition costs, missing parameters, or the inaccessibility of key experimental variables.

Our methodology shows substantial improvements in both prediction accuracy and model scalability. Specifically, the fusion framework elevates titanium alloy ductility prediction accuracy from 0.312 (typical of conventional machine learning) to 0.849. In transfer learning tasks using a diverse dataset of refractory multi-principal-element alloys (RMPEAs), the model achieved a coefficient of determination ($R^2$) of 0.680, demonstrating notable generalization across alloy systems. For 19 out of 25 test samples, the error between predicted and experimental elongation values was less than 5%, exemplifying the method's robustness even without explicit microstructural information. Applying this contrastive learning model to the vast compositional space of the Ti-V-Zr-Nb-Hf-Ta six-component system enables ductility prediction of Ti-containing RMPEAs without requiring prior knowledge of microstructures, successfully identifying alloy systems with superior ductility, such as Ti-V-Ta, Ti-Nb-Ta and Ti-V-Nb-Ta systems, and revealing complex, non-monotonic relationships between composition and ductility that differ fundamentally from empirical rules for conventional alloys. While the model exhibits a tendency to underestimate ductility in more brittle compositions and has limited capacity to capture fine-grained non-monotonic trends under subtle compositional changes, its predictions nonetheless provide a valuable framework for preliminary alloy design and materials property exploration.

Despite these advances, several promising directions remain open for further improvement. First, although our framework shows strong generalization capacity, its performance declines when transferring between alloy systems from titanium alloys to RMPEAs. Future efforts will explore few-shot learning and advanced domain adaptation to further boost transferability and minimize reliance on large labeled datasets. Second, the adoption of reinforcement learning from expert feedback offers opportunities to better align the model's reasoning with materials science expertise, thus improving interpretability and prediction reliability. Together, these directions will help expand the applicability of information fusion models and accelerate the knowledge-driven discovery of advanced materials.

In summary, by efficiently integrating domain-specific knowledge from scientific literature with limited sets of physical descriptors through information fusion and contrastive learning, this work provides a systematic and holistic approach for property prediction in complex material systems, particularly suited for high-dimensional, multi-component alloys and data-driven knowledge discovery scenarios where quantitative descriptors are incomplete or difficult to obtain. Our framework offers a promising approach to address key challenges in materials modeling and provides a foundation for future research in knowledge-guided materials design and informatics-driven materials discovery.

## Materials and Methods

### Dataset construction

We utilized two APIs provided by Elsevier: ScienceDirect Search V2 and Article Retrieval. The ScienceDirect Search V2 API was employed for literature retrieval, allowing us to obtain relevant bibliographic information-including titles, authors, and DOIs-to enhance our corpus and facilitate subsequent traceability. Meanwhile, the Article Retrieval API was used to download full texts in XML format. The search utilized the following keywords: ((Ti alloy) OR (multi-principal-element alloy) OR (high-entropy alloy) AND (mechanical properties) OR (elongation) OR (ductility)). We downloaded the bibliographic information of the top 500 articles, sorted by relevance, and saved this data as a JSON file. Following this, we extracted the DOI numbers from the JSON file and adjusted the parameters of the Article Retrieval API to facilitate literature acquisition based on these DOIs. Ultimately, we successfully saved the full texts in XML format for the top 500 returned articles.

In constructing our corpus, all textual data concerning alloy synthesis pathways and subsequent thermal-mechanical operations were systematically retained. This included procedural details on sample preparation and heat treatments. For microstructure and composition-related sections, narrative statements elucidating the effects of alloying elements or structural characteristics on mechanical behaviors were preserved, even in the absence of specific measurement data. Descriptions solely providing microstructural observations, without comparative context, were also incorporated where relevant. To enhance dataset consistency, any specialized processing techniques documented on a single occasion throughout the corpus were omitted. This refinement reduced the original titanium alloy dataset from 313 entries to 226, which were subsequently allocated into training and validation subsets at an 80:20 ratio. As a result, vacuum arc melting and powder metallurgy emerged as the principal preparation routes.

When evaluating alloys subjected to identical fabrication procedures but exhibiting distinct microstructural profiles and ductility owing to compositional variations, a frequently encountered scenario, the protocol involved selecting only those passages that clearly described unique microstructural features. Alloy identifiers were mapped to vector representations to highlight differences in chemical makeup. Regarding the RMPEA data, 30 representative entries were curated for external validation. In addition, to interrogate the model's capability to discern composition-ductility relationships under fixed processing conditions, 25 entries featuring uniform manufacturing methods and alloy systems but with variant elemental compositions were reserved for in-depth analysis.

## Information fusion model construction

The information fusion model employs a multi-text embedding architecture built upon the domain-adapted pre-trained language model MatSciBERT[41]. By simultaneously encoding the semantic information of material processing methods, chemical compositions, and microstructural descriptions—supplemented by an alloy name prompting strategy—the model is able to accurately predict fracture elongation. The framework consists of two main modules: (1) Text Embedding Extraction, which transforms diverse textual inputs into high-dimensional semantic embeddings; and (2) Downstream Task Module, which combines the fused embeddings for elongation prediction. The models in the present study are trained based on the PyTorch framework using model weights provided by Hugging Face and two NVIDIA Tesla A100 GPUs. To reduce memory overhead and computational time, mixed-precision training is utilized for model optimization. All implementation details are fully documented in our publicly available code repository. The construction and functionality of each module are described as follows.

**Text embedding extraction.** The model processes three types of textual inputs, corresponding to ① materials numerical features into text input via a given template, ②processing details and ③compositional and microstructural characteristics. Each input type is first tokenized using the WordPiece algorithm, a strategy commonly adopted in BERT-based models (Fig. 2a). This approach segments text into subword units, which is particularly advantageous for handling the domain-specific terminology and neologisms prevalent in materials science literature. While tokenization converts raw text into discrete tokens, these lack inherent semantic relationships. Therefore, each token sequence is projected into a continuous vector space via the embedding mechanism, capturing both syntactic and semantic features of the input. Given the technical complexity and unique vocabulary of materials science texts, MatSciBERT is used as the core encoder for this step. As an attentional pre-trained language model specifically trained on a large corpus of materials science documents, MatSciBERT outperforms generic alternatives like BERT and SciBERT, achieving higher F1-scores on materials entity recognition benchmarks. This superior performance ensures that text embeddings reflect the nuanced context and specific content essential for subsequent fracture elongation prediction.

To address embedding ambiguities caused by multi-alloy descriptions within single sentences, we define semantic vector representations as follows, for a given input text, MatSciBERT sequentially generates representative vectors $x_t$ for all tokens:

$$x_t = \mathrm{MatSciBERT}(t) \tag{1}$$

following the training methodology and output format of BERT-based models, the model designates a header vector [CLS] and a tail vector [SEP][58], $x_t$ is formally represented as:

$$x_t = [x_{CLS}, x_{alloy}, x_1, \cdots x_n, x_{SEP}] \tag{2}$$

The $x_{CLS}$ serves as the sentence-level feature, capturing the aggregated semantic information of the entire input sequence, while $x_n$ represents the contextual semantic information of each token. Therefore, for any textual feature $f$, we define its semantic vector representation as:

$$f = x_{cls}, if \quad t = ①$$
$$f = \text{concatenate}(x_{alloy}, x_{CLS}), if \quad t = ②, ③ \tag{3}$$

this encodes the semantic information of the target alloy.

**Downstream task implementation.** To achieve information fusion, as illustrated in Fig. 2b, we concatenate the three text embeddings obtained through the aforementioned approach:

$$F = \text{concatenate}(f_1, f_2, f_3) \tag{4}$$

where $f_1, f_2, f_3$ denote the semantic vectors of the three text features, respectively. We posit that the concatenated embeddings integrate high-dimensional representations of material properties, processing methods, compositions, and microstructures, thereby incorporating significantly more information than traditional machine learning methods that solely rely on material properties as input. Following the acquisition of all textual feature representations, two fully connected layers are employed to predict the alloy's elongation:

$$Elongation = \text{ReLU}(\, W \cdot F + b) \tag{5}$$

where $W$, $b$ denote the trainable parameters (weight matrix and bias vector, respectively) of the fully connected layers, and ReLU serves as the activation function. This design ensures that only the fully connected layer parameters are trainable during downstream task implementation, while all MatSciBERT parameters remain frozen. This decision is motivated by two key considerations: firstly, MatSciBERT exhibits strong representational capacity for materials science texts. secondly, the relatively limited dataset size (226 samples) is insufficient to effectively fine-tune a model with such extensive parameters, which would otherwise incur prohibitive computational costs. To mitigate overfitting, dropout layers and batch normalization are integrated into the fully connected layers. The training objective is to minimize the mean squared error (MSE) between predicted and ground-truth values:

$$MSELoss = \frac{1}{n}\sum_{i=1}^{n}(y_i - \hat{y}_i)^2 \tag{6}$$

where $y_i$ denotes the ground truth value and $\hat{y}_i$ denotes the model's predicted value.

## Contrastive learning model construction

The contrastive learning model proposed in this work establishes a framework for material property prediction through multi-textual domain alignment. By training the model to recognize similarity and dissimilarity among the text embeddings of material properties, processing methods, chemical compositions and microstructures, a representation space is constructed in which the effects of known variables (properties and processing) are disentangled from those of unknown variables (composition and microstructure). This disentanglement allows the model to predict fracture elongation based solely on material numerical features and processing descriptions. The detailed construction methodology is presented in the following sections.

A central aspect of the contrastive learning approach lies in enabling the text embedding extractor to effectively capture and distinguish the relationships among property, processing, composition, and microstructure texts. MatSciBERT, pre-trained specifically on materials science literature, is utilized as the backbone encoder (denoted as "Encoder" in Fig. 2c) for generating these embeddings. To balance the limitations imposed by a relatively small dataset and restricted computational resources, fine-tuning is limited to only the parameters in the 12th layer of MatSciBERT, thus ensuring efficient adaptation while maintaining domain-specific representation capability. Following Equations (1)-(3) from the information fusion model, the three independent Encoders generate corresponding text vectors $f_1, f_2, f_3$ for given text inputs. According to the determinability of alloy text features prior to experimentation, we design the text embeddings as follows:

$$I_1 = W_p(\text{concatenate}(f_1, f_2)) + b_p$$
$$I_2 = f_3 \quad (7)$$

where $W_p$ and $b_p$ denote the weight matrix and bias vector, respectively, of the projection layer applied after concatenating the processing and properties text embeddings. This transformation ensures dimensional consistency between the resulting vectors $I_1$ and $I_2$.

This work adopts the InfoNCE loss - a contrastive learning objective used in both MOCO[59] and CLIP[60] to train the model. The optimization goal is to minimize the cosine similarity between positive and negative sample pairs:

$$L_q = -\log \frac{\exp(I_1 \cdot I_2/\tau)}{\sum_{i=0}^{K} \exp(I_1 \cdot I_2/\tau)} \quad (8)$$

where K denotes the batch size, which determines the number of positive-negative sample pairs per batch. In contrastive learning, larger batch sizes provide more negative samples for comparison with each positive sample, thereby enhancing the model's discriminative capability-at the cost of increased computational resources. Due to hardware and dataset constraints (226 samples), we set K=128. The temperature coefficient $\tau$ controls the sharpness of the similarity distribution. This tunable hyperparameter is optimized via random search in our experiments.

Upon completing contrastive learning training, we freeze all parameters of both the ProcsEncoder and PropsEncoder, rendering them non-trainable. Following Equations (1)-(3) in the information fusion model, we obtain the combined text embedding $I_1$:

$$I_1 = \text{concatenate}(\text{PropsEncoder}(t^{①}) + \text{ProcEncoder}(t^{②})) \tag{9}$$

where PropsEncoder and ProcEncoder denote the material properties text encoder and processing methods text encoder, respectively, while $t^{①}$ and $t^{②}$ denote the material property text and processing method text inputs. Finally, we adopt the downstream task implementation of the information fusion model to predict alloy elongation:

$$Elongation = \text{Relu}(W \cdot I_1 + b) \tag{10}$$

and update the parameters $W$ and $b$ using the MSE loss function (as shown in Equation (6). Dropout layers and batch normalization are also added to prevent overfitting.

## Materials numerical features model construction

Three classical statistical machine learning algorithms—Support Vector Machines (SVM), Random Forests (RF), and Multilayer Perceptron (MLP)—are utilized to predict alloy fracture elongation using our curated dataset. 18 numerical physical parameters serve as input features, as defined in the main text. Prior to model training, all features are normalized, and the dataset is randomly partitioned into training (80%) and testing (20%) subsets. Model evaluation is based on the $R^2$, with its mathematical equations given later. Key hyperparameters for each algorithm are optimized via grid search over predefined ranges, employing 5-fold cross-validation on the training set and using $R^2$ as the selection metric. The best-performing configurations and their associated hyperparameters (all other parameters retained default values) are summarized in Table S1. All machine learning workflows and implementations are carried out using the scikit-learn library.

## Direct fusion model of textual and materials numerical features

In the direct fusion strategy, the textual features extracted by MatSciBERT are concatenated with the normalized values of 18 materials numerical features to construct a combined vector of dimension 768×4+18. This combined vector is then subjected to L2 normalization before being fed into a neural network composed of two hidden layers, with layer sizes (768×4+18, 768) and (768, 1), respectively. The dataset is partitioned in accordance with the procedure described in the Datasets section of the Results. Fracture elongation serves as the prediction target (label), and the network is trained using the mean squared error loss (MSELoss) function. The optimization is performed using Adam with a learning rate of 1e-3, over 120 epochs, batch size of 32, and a weight decay rate of 1e-4 to mitigate overfitting. The model's fitting performance is illustrated in Fig. S1a.

## Fusion of transformed textual and materials numerical features

In contrast to the direct fusion strategy, this approach first transforms the textual and physical numerical features separately using two fully connected (Linear) layers. The textual features output by MatSciBERT are passed through a linear layer (768×4), while the 18 physical numerical features are transformed via a linear layer with dimensions 18. The outputs from these two layers are then concatenated to form a new composite vector, which is subsequently used as the input for the same neural network architecture as in the direct fusion strategy. All hyperparameter settings remain identical. The model's performance for this strategy is presented in Fig. S1b.

## GloVe-based information fusion model construction

GloVe (Global Vectors for Word Representation) is a word embedding model that learns vector representations by analyzing word co-occurrence statistics in large text corpora, capturing semantic properties such as similarity and analogy between words. We employed GloVe for word vector acquisition and fracture elongation prediction following these steps: First, we loaded the pre-trained GloVe-6B-100d model, which provides 100-dimensional vectors trained on 6 billion tokens. Second, we mapped preprocessed text words to the GloVe vocabulary and obtained corresponding word vectors through three nn.Embedding layers for the three text types.

Third, we concatenated word vectors from all three text types to achieve semantic aggregation, then used a bidirectional GRU layer to generate sentence vectors. The GRU was selected for its ability to capture long- and short-term dependencies with fewer parameters than LSTM, enabling faster convergence. Finally, sentence vectors were fed through two linear layers (200→100→1 nodes) to predict fracture elongation. The model was trained using MSE loss with Adam optimizer (learning rate: 1e-3, 300 epochs, batch size: 64). Weight decay (1e-3) and dropout (0.5) were applied to prevent overfitting.

## Question-answering model based on DeepSeek-V3

DeepSeek-V3 is a large-scale pretrained language model developed independently by the DeepSeek team. Based on the Transformer architecture, it is trained using self-supervised learning on massive textual corpora. The model is capable of effectively capturing deep semantic and syntactic features and supports a wide range of natural language processing tasks, including text generation, question answering, machine translation, and code generation. As one of the state-of-the-art large language models, DeepSeek-V3 demonstrates robust capabilities in both general and domain-specific language understanding and generation.

Under zero-shot conditions, we utilized DeepSeek-V3 to predict the fracture elongation of alloys. Specifically, we accessed the DeepSeek model via its API and input textual descriptions of alloy material properties, processing methods, and composition/microstructure information as prompts to query the model for the target alloy's ductility. Predictions were performed separately for the titanium alloy dataset and the RMPEA dataset. Subsequently, the performance of DeepSeek's predictions was quantitatively evaluated using the $R^2$.

## Evaluation methods

The two regression evaluation metrics employed in this study are the Root Mean Square Error (RMSE) and the coefficient of determination ($R^2$), whose computational formulas are shown in Equations (11) and (12), respectively. In these equations, $y_i$ represents the observed values in the dataset, $\bar{y}$ denotes the mean of all observed values, and $\hat{y}_i$denotes the model's predicted values. RMSE quantifies the deviation between predicted and actual values, with smaller values indicating better predictive accuracy. $R^2$ measures the overall goodness-of-fit of the model, where values closer to 1 represent superior fitting performance.

$$RMSE = \sqrt{\frac{1}{n}\sum_{i=1}^{n}(y_i - \hat{y}_i)^2} \tag{11}$$

$$R^2 = 1 - \frac{\sum_{i=1}^{n}(y_i - \hat{y}_i)^2}{\sum_{i=1}^{n}(y_i - \bar{y})^2} \tag{12}$$

## Data availability

All the training data and test data will be released upon publication.

## Code availability

The code in this work will be released upon publication.

## Acknowledgments

This work is financially supported by National Natural Science Foundation of China [U20A20231].

## Author contributions

YP and ZZ conducted the investigation, formal analysis, data curation, methodology development, and wrote the original draft and reviewed the manuscript. LZ performed data curation, and formal analysis. FY and YL contributed to the corpus construction and model construction. YY provided project administration and interpreted the results. SB secured funding acquisition and contributed to project administration. All authors read and approved the final manuscript.

## Competing interests

The authors declare that they have no known competing financial interests or personal relationships that could have appeared to influence the work reported in this paper.

## Reference

1. Sanchez-L, B. *et al.* A. Inverse molecular design using machine learning: Generative models for matter engineering. *Science* **361**, 360–365 (2018).
2. Butler, K. T. *et al.* Cartwright, H., Isayev, O. & Walsh, A. Machine learning for molecular and materials science. *Nature* **559**, 547–555 (2018).
3. Xu, P. *et al*. Small data machine learning in materials science. *NPJ Comput. Mater.* **9**, 42 (2023).
4. Szymanski, N. J. *et al.* An autonomous laboratory for the accelerated synthesis of novel

materials. *Nature* **624**, 86–91 (2023).
5. Zeni, C. *et al.* A generative model for inorganic materials design. *Nature* **639**, 624–632 (2025).
6. Chang, H. *et al*. Phase prediction and effect of intrinsic residual strain on phase stability in high-entropy alloys with machine learning. *J. Alloys Compd.* **921**, 166149 (2022).
7. Jacobs, R. *et al.* Machine learning materials properties with accurate predictions, uncertainty estimates, domain guidance, and persistent online accessibility. *Mach. Learn.-Sci. Technol.* **5**, 045051 (2024).
8. Mahadevkar, S. *et al.* A Review on Machine Learning Styles in Computer Vision-Techniques and Future Directions. *IEEE Access* **10**, 107293–107329 (2022).
9. Zhang, H. *et al*. Revealing the technology development of natural language processing: A Scientific entity-centric perspective. *Inf. Process. Manag.* **61**, 103574 (2024).
10. Ye, Y. *et al.* Improving machine learning based phase and hardness prediction of high-entropy alloys by using Gaussian noise augmented data. *Comput. Mater. Sci.* **223**, 112140 (2023).
11. Sun, Y. *et al.* EFTGAN: Elemental features and transferring corrected data augmentation for the study of high-entropy alloys. *npj Comput. Mater.* **11**, 54 (2025).
12. Rao, Z. *et al.* Machine learning–enabled high-entropy alloy discovery. *Science* **378**, 78–85 (2022).
13. Lookman, T. *et al*. Active learning in materials science with emphasis on adaptive sampling using uncertainties for targeted design. *npj Comput. Mater.* **5**, 21 (2019).
14. Jiang, L. *et al.* A rapid and effective method for alloy materials design via sample data transfer machine learning. *npj Comput. Mater.* **9**, 26 (2023).
15. Gupta, V. *et al.* Structure-aware graph neural network based deep transfer learning framework for enhanced predictive analytics on diverse materials datasets. *npj Comput. Mater.* **10**, 1 (2024).
16. Yang, C. *et al.* A machine learning-based alloy design system to facilitate the rational design of high entropy alloys with enhanced hardness. *Acta Mater.* **222**, 117431 (2022).
17. Tran, H. D. *et al.* Machine-learning predictions of polymer properties with Polymer Genome. *J. Appl. Phys.* **128**, 171104 (2020).
18. Sathiyamoorthi, P. & Kim, H. S. High-entropy alloys with heterogeneous microstructure: Processing and mechanical properties. *Prog. Mater. Sci.* **123**, 100709 (2022).
19. Nutor, R. K. *et al.* A dual-phase alloy with ultrahigh strength-ductility synergy over a wide temperature range. *Sci. Adv.* **7**, eabi4404 (2021).
20. Xiong, W. *et al*. Refractory high-entropy alloys: A focused review of preparation methods and properties. *J. Mater. Sci. Technol.* **142**, 196–215 (2023).
21. Sathiyamoorthi, P. & Kim, H. S. High-entropy alloys with heterogeneous microstructure: Processing and mechanical properties. *Prog. Mater. Sci.* **123**, 100709 (2022).
22. Amiri, M. *et al*. Unveiling processing-property relationships in laser powder bed fusion: The synergy of machine learning and high-throughput experiments. *Mater. Des.* **252**, 113705 (2025).
23. Hamid, M. *et al*. A dislocation-based stress-strain gradient plasticity model for strength and ductility in materials with gradient microstructures. *Philos. Mag.* **98**, 2896–2916 (2018).
24. Cao, Y. *et al.* Machine learning assisted prediction and optimization of mechanical

properties for laser powder bed fusion of Ti6Al4V alloy. *Addit. Manuf.* **91**, 104341 (2024).
25. Wang, J. *et al.* High strength-ductility Co23Cr23Ni23Mn31 medium-entropy alloy achieved via defect engineering. *Mater. Sci. Eng. A* **796**, 139974 (2020).
26. Hart, G. L. W. *et al*. Machine learning for alloys. *Nat. Rev. Mater.* **6**, 730–755 (2021).
27. Mehranpour, M. S. *et al*. Machine learning prediction and characterization of sigma-free high-entropy alloys. *Mater. Charact.* **212**, 113937 (2024).
28. Bundela, A. S. & Rahul, M. R. Machine learning-enabled framework for the prediction of mechanical properties in new high entropy alloys. *J. Alloys Compd.* **908**, 164578 (2022).
29. Zhao, F. *et al.* Machine learning guided prediction of dynamic energy release in high-entropy alloys. *Mater. Des.* **246**, 113339 (2024).
30. Wang, W. *et al.* Alloy synthesis and processing by semi-supervised text mining. *npj Comput. Mater.* **9**, 183 (2023).
31. Negahdary, M. & Mabbott, S. Automated synthesis and processing of functional nanomaterials: Advances and perspectives. *Coord. Chem. Rev.* **523**, 216249 (2025).
32. Kononova, O. *et al.* Text-mined dataset of inorganic materials synthesis recipes. *Sci. Data* **6**, 203 (2019).
33. Vaswani, A. *et al.* Attention is All you Need. in *Advances in Neural Information Processing Systems*. **30** (2017).
34. Dagdelen, J. *et al.* Structured information extraction from scientific text with large language models. *Nat. Commun.* **15**, 1418 (2024).
35. Gupta, S. *et al*. Data extraction from polymer literature using large language models. *Commun. Mater.* **5**, 1–11 (2024).
36. Polak, M. P. & Morgan, D. Extracting accurate materials data from research papers with conversational language models and prompt engineering. *Nat. Commun.* **15**, 1569 (2024).
37. Ye, Y. *et al.* Construction and Application of Materials Knowledge Graph in Multidisciplinary Materials Science via Large Language Model. *Adv. Neural Inf. Process. Syst.* **37**, 56878–56897 (2024).
38. Bai, X. *et al.* Construction of a knowledge graph for framework material enabled by large language models and its application. *npj Comput. Mater.* **11**, 51 (2025).
39. Tshitoyan, V. *et al.* Unsupervised word embeddings capture latent knowledge from materials science literature. *Nature* **571**, 95–98 (2019).
40. Sasidhar, K. N. *et al.* Enhancing corrosion-resistant alloy design through natural language processing and deep learning. *Sci. Adv.* **9**, eadg7992 (2023).
41. Gupta, T. *et al*. MatSciBERT: A materials domain language model for text mining and information extraction. *npj Comput. Mater.* **8**, 102 (2022).
42. Shetty, P. *et al.* A general-purpose material property data extraction pipeline from large polymer corpora using natural language processing. *npj Comput. Mater.* **9**, 52 (2023).
43. Kuenneth, C. & Ramprasad, R. polyBERT: a chemical language model to enable fully machine-driven ultrafast polymer informatics. *Nat. Commun.* **14**, (2023).
44. Tian, S. *et al.* Steel design based on a large language model. *Acta Mater.* **285**, 120663 (2025).
45. Yang, C. *et al.* A machine learning-based alloy design system to facilitate the rational design of high entropy alloys with enhanced hardness. *Acta Mater.* **222**, 117431 (2022).
46. Bundela, A. S. & Rahul, M. R. Machine learning-enabled framework for the prediction of

mechanical properties in new high entropy alloys. *J. Alloys Compd.* **908**, 164578 (2022).
47. Tao, Q. *et al.* Machine learning strategies for small sample size in materials science. *Sci. CHINA-Mater.* **68**, 387–405 (2025).
48. Pennington, J., Socher, R. & Manning, C. GloVe: Global Vectors for Word Representation. in *Proceedings of the 2014 conference on empirical methods in natural language processing (EMNLP).* 1532–1543 (2014).
49. DeepSeek-AI *et al.* DeepSeek-V3 Technical Report. Preprint at https://doi.org/10.48550/arXiv.2412.19437 (2025).
50. Mirza, A. *et al.* A framework for evaluating the chemical knowledge and reasoning abilities of large language models against the expertise of chemists. *Nat. Chem.* 1–8 (2025).
51. McDuff, D. *et al.* Towards accurate differential diagnosis with large language models. *Nature* 1–7 (2025).
52. Huang, H. *et al.* Phase-Transformation Ductilization of Brittle High-Entropy Alloys via Metastability Engineering. *Adv. Mater.* **29**, 1701678 (2017).
53. Yuan, Y. *et al.* Formation, structure and properties of biocompatible TiZrHfNbTa high-entropy alloys. *Mater. Res. Lett.* **7**, 225–231 (2019).
54. Zýka, J. *et al.* Microstructure and Room Temperature Mechanical Properties of Different 3 and 4 Element Medium Entropy Alloys from HfNbTaTiZr System. *Entropy* **21**, 114 (2019).
55. Ozan, S. *et al*. Development of Ti–Nb–Zr alloys with high elastic admissible strain for temporary orthopedic devices. *Acta Biomater.* **20**, 176–187 (2015).
56. Nguyen, V. T. *et al.* Compositional design of strong and ductile (tensile) Ti-Zr-Nb-Ta medium entropy alloys (MEAs) using the atomic mismatch approach. *Mater. Sci. Eng. A* **742**, 762–772 (2019).
57. Dirras, G. *et al.* Elastic and plastic properties of as-cast equimolar TiHfZrTaNb high-entropy alloy. *Mater. Sci. Eng. A* **654**, 30–38 (2016).
58. Devlin, J., Chang, M.-W. & Lee, K. BERT: Pre-training of Deep Bidirectional Transformers for Language Understanding. in *Proceedings of the 2019 Conference of the North American Chapter of the Association for Computational Linguistics: Human Language Technologies*. **1**, 4171–4186 (2019).
59. He, K. *et al*. Momentum Contrast for Unsupervised Visual Representation Learning. in *2020 IEEE/CVF Conference on Computer Vision and Pattern Recognition (CVPR)*. 9726–9735 (2020).
60. Radford, A. *et al.* Learning Transferable Visual Models From Natural Language Supervision. in *Proceedings of the 38th International Conference on Machine Learning* (PmLR). 8748–8763 (2021).

# Supplementary Information

# for

# Information fusion strategy integrating pre-trained language model and contrastive learning for materials knowledge mining

Yongqian Peng[+], Zhouran Zhang[+,*], Longhui Zhang, Fengyuan Zhao, Yahao Li, Yicong Ye[*], Shuxin Bai

+These authors contributed equally to this work. *Corresponding authors.

College of Aerospace Science and Engineering, National University of Defense Technology

Changsha, Hunan, China, 410073

## Materials numerical features models

Table S1 The best-performing configurations and their associated hyperparameters as well as the $R^2$ and RMSE of SVM, RF and MLP models

| Model | hyperparameter | $R^2$/RMSE | | |
|---|---|---|---|---|
| | | Train | Validation | Test |
| SVM | gamma = 10,<br>c = 10; | 0.264/9.77 | 0.400/10.34 | -0.361/13.61 |
| RF | n_estimators = 200,<br>max_features = 2; | 0.331/9.21 | 0.476/9.67 | 0.135/10.85 |
| MLP | hidden_layer_sizes = (30, 30),<br>learning_rate_init = 1e-3 | 0.175/10.27 | 0.312/11.08 | -0.903/16.09 |

## Numerical-text model and numerical-text embedded model

In the dataset constructed for this study, three primary types of input are included: (1) numerical features derived from 18 physical parameters, (2) textual descriptions of alloy preparation and heat treatment processes, and (3) textual descriptions of compositional variations and microstructures. Effectively integrating these three categories of features is one of the key aspects in building the information fusion model. In addition to the fusion strategies highlighted in the main text, we also explored two alternative approaches for integrating these feature types, which are direct fusion model of textual and materials numerical features (numerical-text model) and fusion of transformed textual and materials numerical features (numerical-text embedded model). The fitting performance of the direct fusion model—where textual features and materials numerical features are directly combined—is shown in Fig. S1(a). The fitting results for the strategy that combines transformed textual features with materials numerical features are presented in Fig. S1(b). For textual feature extraction, as described in the main text, raw text cannot be directly input into machine learning models and must be transformed into numerical representations via tokenization and word embedding. However, these initial steps only provide sparse vector representations without capturing semantic relations, limiting their effectiveness in conveying meaning. Because MatSciBERT is pretrained on large-scale general, scientific, and materials science corpora, it exhibits strong capability in extracting salient features from materials science texts. Therefore, the pretrained language model MatSciBERT is also employed in these two information fusion strategies to extract textual features. The extraction process follows the methodology described in the main text, using the CLS vector generated by MatSciBERT as the sentence representation and concatenating it with the alloy token embeddings to form a new semantic vector. This combined vector, with a dimensionality of 768×4 (sentence and token embeddings each have a dimension of 768, for both processing-related and composition/microstructure-related texts), is used as the extracted textual feature in subsequent modeling.

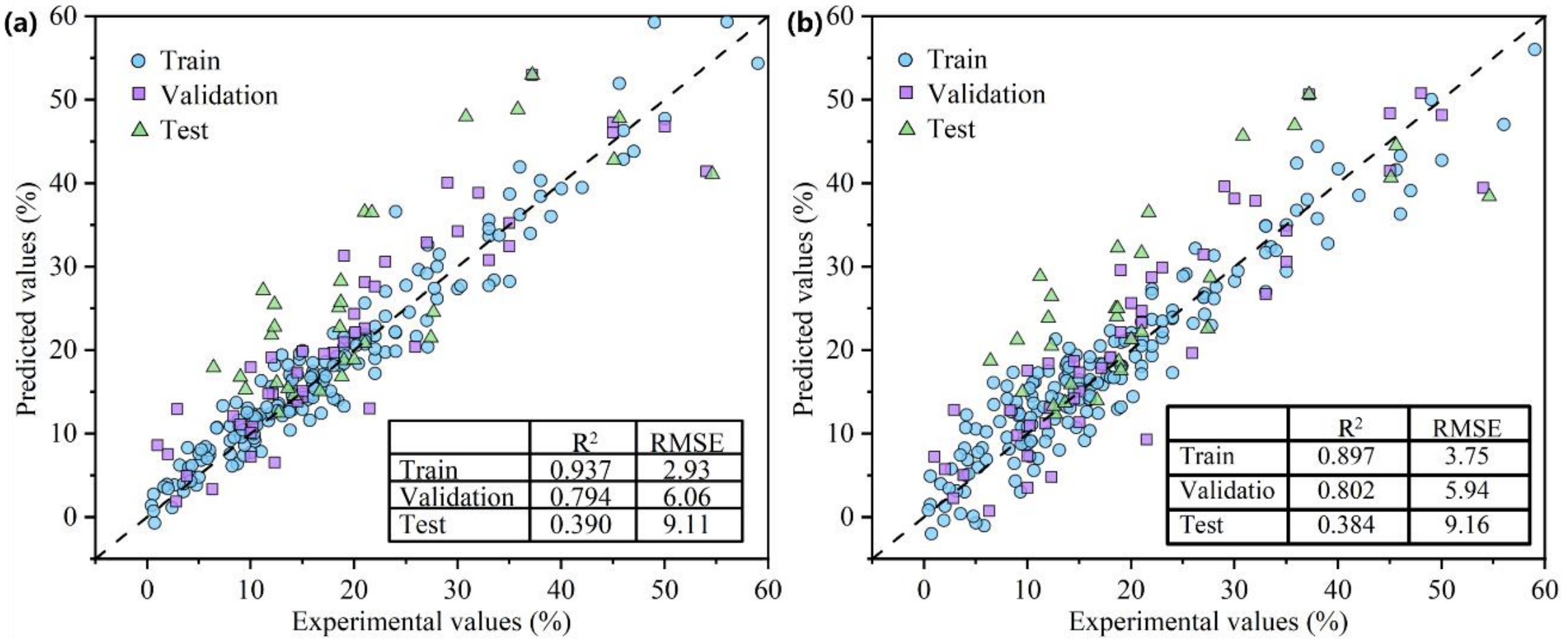

Fig. S1. Fitting performance of models constructed using different fusion strategies on the dataset: (a) direct fusion of textual features and materials numerical features; (b) fusion of transformed textual features and materials numerical features.

## GloVe-based information fusion model

In addition to MatSciBERT, we also employed GloVe for word vector acquisition and fracture elongation prediction in order to compare the semantic understanding and representation capabilities of different models. The loss curve of the GloVe-based information fusion model is shown in Fig. S2(a). Despite the application of regularization, the loss on the test set remains substantially higher than that on the training set, indicating that the GloVe-based model exhibits a certain degree of overfitting. The model's fitting performance is presented in Fig. S2(b).

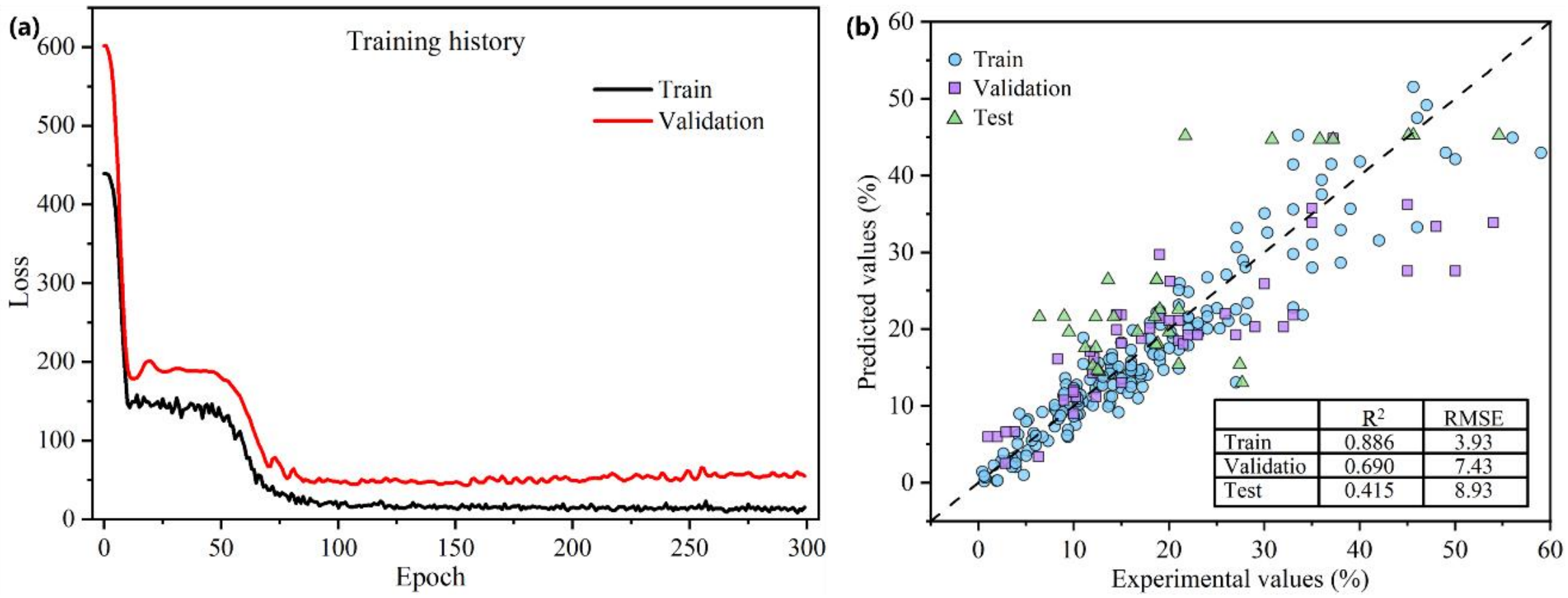

Fig. S2. Training results of the GloVe-based information fusion model. (a) Loss curves of the model, illustrating the changes in mean squared error (MSE) on the training and test sets during training. (b) Fitting performance of the model on the dataset; the scattered distribution of data points around the diagonal indicates poor fitting performance on the test set.

## Question-answering model based on DeepSeek-V3

Table S2 Example of alloy fracture elongation Q&A records based on DeepSeek V3

| |
|---|
| **Alloy #1** |
| **Prompt:**<br>Hello, assuming you are a materials science researcher, given the properties of the alloy, the processing method of the alloy, and the composition and microstructure of the alloy. Please predict the fracture elongation of the target alloy for me, remembering that the target alloy is the second word of the alloy's properties, or the first word of the alloy's process method, composition, and microstructure text. |
| **Answer:**<br>To predict the fracture elongation of the target alloy, I'll need you to provide the following details: Properties of the alloy (the target alloy is the second word of this text). Process method of the alloy (the target alloy is the first word of this text). Composition and microstructure of the alloy (the target alloy is the first word of this text). Once you supply this information, I can analyze the data and predict the fracture elongation. For example: If the properties are "high strength aluminum," the target alloy is "strength." If the process method is "rolled steel," the target alloy is "rolled." If the composition is "titanium-based composite," the target alloy is "titanium-based." Please provide the relevant details, and I'll proceed with the prediction! |
| **Prompt:**<br>For Ti-12Mo alloy, the average atomic radius is 1.319, the atomic radius mismatch is 0.00371, the atomic radius variance is 0.00490, the average electronegativity is 1.580, the electronegativity mismatch is 0.09606, the electronegativity variance is 0.15175, the average valence electron concentration is 4.128, the valence electron concentration mismatch is 0.11858, the valence electron concentration variance is 0.48951, the average Young's modulus is 129.632, the Young's modulus mismatch is 0.40216, the Young's modulus variance is 52.13238, the shear modulus is 49.248, the shear modulus mismatch is 0.40752, the shear modulus variance is 20.06974, the bulk modulus is 117.680, the bulk modulus mismatch is 0.24958, the bulk modulus variance is 29.37035.<br>Ti-12Mo. The ternary Ti–12Mo-5Zr (wt%) alloy was fabricated by the arcmelting furnace under pure Ar atmosphere using pure titanium, molybdenum and zirconium. The alloy ingot about 3 kg was melted for five times, and flipped over each time before melting. Then, the ingot was homogenized at 1173 K for 72 ks under high vacuum of 10−4 Pa, followed by water quenching. The as-quenched ingot was heavily cold rolled to 0.5 mm in thickness at a reduction rate of > 95% at room temperature. From the as-rolled sheet, tensile specimens were mechanically prepared with gauge dimensions of 60 mm (length) × 5 mm (width) × 0.5 mm (thickness). The specimens were solution-treated (ST) at 1173 K for 1.8 ks under high vacuum atmosphere and subsequently water-quenched.<br>Ti-12Mo. Compared with Ti-12Mo alloy, the Ti-12Mo-5Zr alloy is improved by 30% in keeping almost the same uniform ductility, suggesting that the zirconium addition may act as a solution strengthening alloying element. |
| **Answer:**<br>Fracture Elongation Prediction for Ti-12Mo Alloy<br>Given the provided data, we can analyze the key factors influencing fracture elongation (ductility) in the Ti-12Mo alloy:<br>1. Key Parameters Affecting Ductility<br>Atomic radius mismatch (0.00371) → Low, suggesting minimal lattice distortion, which is good for ductility.<br>Young's modulus (129.632 GPa) & Shear modulus (49.248 GPa) → Moderate, not too stiff, allowing some plastic deformation.<br>Cold rolling (95% reduction) → Severe deformation typically reduces ductility unless followed by recrystallization.<br>Solution treatment (ST at 1173 K for 1.8 ks) → Helps recover ductility by removing cold-worked |

microstructure.

Zr addition (in Ti-12Mo-5Zr) → Improves strength while maintaining ductility (reference: 30% improvement).

2. Comparison with Similar Alloys

Pure Ti → ~20-30% elongation.

Ti-6Al-4V → ~10-15% elongation (strong but less ductile).

β -Ti alloys (like Ti-12Mo) → Typically 10-20% elongation, depending on processing.

Ti-12Mo-5Zr reference → "Almost the same uniform ductility" suggests Ti-12Mo is comparable (~15-20%).

3. Predicted Fracture Elongation for Ti-12Mo

Considering:

Moderate mismatch values (favoring ductility).

Cold rolling + solution treatment (balancing strength and ductility).

Similarity to Ti-12Mo-5Zr (~15-20% elongation).

Final Prediction:

Fracture elongation of Ti-12Mo ≈ 15–20% (likely ~18% under optimal ST conditions).

**Prompt:**

When you answer me, just go straight to the third step and provide the Key Justifications

Do not provide a range, please give me a value that you are most certain of.

**Answer:**

Predicted Fracture Elongation of Ti-12Mo Alloy: 18%

Key Justifications:

Low atomic radius mismatch (0.00371) → Minimal lattice distortion, supporting ductility.

Cold rolling + solution treatment → Balances work hardening and recrystallization, typical of β-Ti alloys.

Comparison to Ti-12Mo-5Zr reference → Similar uniform ductility suggests Ti-12Mo achieves comparable elongation.

Moderate Young's/shear moduli → Allows plastic deformation without excessive brittleness.

**Alloy #2**

**Prompt:**

for Ti-12Mo-5Zr alloy, the average atomic radius is 1.322, the atomic radius mismatch is 0.01708, the atomic radius variance is 0.02258, the average electronegativity is 1.574, the electronegativity mismatch is 0.10063, the electronegativity variance is 0.15841, the average valence electron concentration is 4.130, the valence electron concentration mismatch is 0.11938, the valence electron concentration variance is 0.49305, the average Young's modulus is 129.033, the Young's modulus mismatch is 0.41023, the Young's modulus variance is 52.93262, the shear modulus is 49.011, the shear modulus mismatch is 0.41588, the shear modulus variance is 20.38281, the bulk modulus is 117.252, the bulk modulus mismatch is 0.25497, the bulk modulus variance is 29.89597.

Ti-12Mo-5Zr. The ternary Ti–12Mo-5Zr (wt%) alloy was fabricated by the arcmelting furnace under pure Ar atmosphere using pure titanium, molybdenum and zirconium. The alloy ingot about 3 kg was melted for five times, and flipped over each time before melting. Then, the ingot was homogenized at 1173 K for 72 ks under high vacuum of 10−4 Pa, followed by water quenching. The as-quenched ingot was heavily cold rolled to 0.5 mm in thickness at a reduction rate of > 95% at room temperature. From the as-rolled sheet, tensile specimens were mechanically prepared with gauge dimensions of 60 mm (length) × 5 mm (width) × 0.5 mm (thickness). The specimens were solution-treated (ST) at 1173 K for 1.8 ks under high vacuum atmosphere and subsequently water-quenched.

Ti-12Mo-5Zr. The X-ray diffraction pattern of Ti-12Mo-5Zr specimens in “as-rolled” and “solution-treated (ST)” states. The XRD pattern obtained on the “as-rolled” specimen shows clearly a dual phase β + α″ microstructure, where the broad β peaks indicate heavily distorted β grains and the strong α″ peaks are referred to the relatively high volume fraction of α″ phase in the β matrix. From the ST specimen, minor athermal ω (ωath) peaks were detected besides the β phase, meaning ωath nanoparticles precipitated during the quench process. It should be mentioned that the α″ phase could not be detected on the ST specimens, which means that the martensite start (Ms) temperature is below room temperature for the Ti-12Mo-5Zr alloy composition. Compared with Ti-12Mo alloy, the Ti-12Mo-5Zr alloy is improved by 30% in keeping almost the same uniform ductility, suggesting that the zirconium addition may act as a solution strengthening alloying element.

**Answer:**

Predicted Fracture Elongation of Ti-12Mo-5Zr Alloy: 22

Key Justifications:

Zr addition (5 wt%) → Solid-solution strengthening without significant ductility loss (reference: 30% improvement over Ti-12Mo while maintaining uniform elongation)

Solution treatment (ST at 1173 K) → Eliminates cold-rolling-induced α″ martensite, retaining ductile β-phase with minor ωath precipitates (minimal embrittlement).

Low atomic radius mismatch (0.01708) and moderate modulus values → Balanced strain

| |
|---|
| accommodation, favoring plasticity.<br>XRD evidence → Suppression of α″ post-ST avoids martensitic brittleness, enhancing ductility compared to as-rolled state. |